\documentclass[aps,twocolumn,prb]{revtex4-1}
\usepackage{epsfig,graphicx}
\begin{document}
\title{Nonlocal van der Waals functionals: The case of rare-gas dimers and solids}
\author{Fabien Tran}
\author{J\"{u}rg Hutter}
\affiliation{Institute of Physical Chemistry, University of Zurich,
Winterthurerstrasse 190, CH-8057 Zurich, Switzerland}

\begin{abstract}

Recently, the nonlocal van der Waals (vdW) density functionals
[M. Dion, H. Rydberg, E. Schr\"{o}der, D. C. Langreth, and B. I. Lundqvist,
Phys. Rev. Lett. \textbf{92}, 246401 (2004)] have
attracted considerable attention due to their good performance for systems
where weak interactions are important. Since the physics of dispersion is
included in these functionals, they are usually more accurate and show
less erratic behavior than the semilocal and hybrid methods.
In this work, several variants of the vdW
functionals have been tested on rare-gas dimers (from He$_{2}$ to Kr$_{2}$)
and solids (Ne, Ar, and Kr) and their accuracy compared to standard
semilocal approximations supplemented or not by an atom-pairwise dispersion
correction [S. Grimme, J. Antony, S. Ehrlich, and H. Krieg,
J. Chem. Phys. \textbf{132}, 154104 (2010)].
An analysis of the results in terms of energy decomposition is also provided.

\end{abstract}

\maketitle

\section{\label{introduction}Introduction}

Thanks to its relatively low cost/accuracy ratio, the Kohn-Sham (KS)
\cite{KohnPR65} version of density functional theory (DFT) \cite{HohenbergPR64}
is the most used quantum method for the calculation of the geometrical and
electronic properties of molecules, surfaces, and solids. The accuracy of the
results of a KS-DFT calculation depends primarily on the chosen approximation
for the exchange-correlation functional (xc) $E_{\text{xc}}$ (see
Ref. \onlinecite{CohenCR12} for a recent review). Nowadays, the most popular
types of approximations for $E_{\text{xc}}$ are the semilocal [in particular
the generalized gradient approximation (GGA) \cite{BeckePRA88,PerdewPRL96}] and
hybrid functionals,\cite{BeckeJCP93} which very often give satisfactory
results. However, it is well known that by construction none of these two
approximations account properly for the dispersion interactions, which arise
due to the attraction between non-permanent dipoles, and that the results
obtained with semilocal and hybrid methods on systems where dispersion
interactions play a major role are often unreliable (see, e.g.,
Refs. \onlinecite{KristyanCPL94} and \onlinecite{PerezJordaCPL95}).

Therefore, efforts have been made to propose methods within the framework of
KS-DFT which explicitly account for the dispersion interactions
(see Refs. \onlinecite{JohnsonJPOC09,TkatchenkoMRSB10,GrimmeWCMS11,KlimesJCP12}
for reviews). Among these methods, the simplest consist of adding to the KS-DFT
total energy a dispersion term of the form
\begin{equation}
E_{\text{disp}} = -\sum_{A<B}\sum_{n=6,8,10,\ldots}
f_{n}^{\text{damp}}(R_{AB})\frac{C_{n}^{AB}}{R_{AB}^{n}},
\label{Edisp}
\end{equation}
where $C_{n}^{AB}$ are the dispersion coefficients for the atom pair $A$ and
$B$ separated by the distance $R_{AB}$ and $f_{n}^{\text{damp}}$ is a damping
function preventing Eq. (\ref{Edisp}) to become too large at small $R_{AB}$.
The coefficients $C_{n}^{AB}$ can be either precomputed (see, e.g., Refs.
\onlinecite{WuJCP02,GrimmeJCC04,GrimmeJCC06}) or calculated using properties
(e.g., electron density) of the system under consideration like in the
exchange-hole dipole moment model (XDM) of Becke and Johnson\cite{BeckeJCP05}
or the method of Tkatchenko and Scheffler.\cite{TkatchenkoPRL09}
The DFT-D2\cite{GrimmeJCC06} and DFT-D3\cite{GrimmeJCP10}
versions of Grimme are currently the most widely used of these methods.
One of the advantages of most methods using Eq. (\ref{Edisp}) is to add
a relatively negligible computational cost compared to the calculation of the
KS-DFT energy. 

Another group of methods accounting explicitly of dispersion interactions
consist of adding a nonlocal term of the form
\begin{equation}
E_{\text{c}}^{\text{nl}} = \frac{1}{2}\int\int\rho(\textbf{r})
\Phi\left(\textbf{r},\textbf{r}'\right)\rho(\textbf{r}')
d^{3}rd^{3}r'
\label{Ecnl}
\end{equation}
to a LDA (local density approximation) or GGA correlation functional.
In Eq. (\ref{Ecnl}), the kernel $\Phi$ depends on quantities at $\textbf{r}$
and $\textbf{r}'$:
\begin{equation}
\Phi\left(\textbf{r},\textbf{r}'\right) =
\Phi\left(\rho(\textbf{r}),\rho(\textbf{r}'),
\left\vert\nabla\rho(\textbf{r})\right\vert,
\left\vert\nabla\rho(\textbf{r}')\right\vert,
\left\vert\textbf{r}-\textbf{r}'\right\vert\right).
\label{Phi}
\end{equation}
The first functional of the form given by Eq. (\ref{Ecnl}), which could be
applied to any type of systems was proposed by Dion \textit{et al}. (DRSLL)
.\cite{DionPRL04}
The DRSLL term was derived starting from the adiabatic connection-fluctuation
dissipation theorem.\cite{LangrethSSC75,GunnarssonPRB76,LangrethPRB77}
Originally, it was used in
combination with revPBE\cite{ZhangPRL98} (a reparametrization of the
Perdew-Burke-Ernzerhof functional PBE\cite{PerdewPRL96})
for exchange and LDA for correlation
and the functional is named as vdW-DF in the literature
(in Table \ref{table1} the composition of the functionals tested in the
present work are given).
vdW-DF was shown to be a clear improvement over the commonly used functionals,
however it became also obvious that a serious shortcoming of vdW-DF is to
systematically overestimate the equilibrium distances.
\cite{LangrethJPCM09,KlimesPRB11}

Therefore, several attempts have been made to
remedy this problem by combining the DRSLL nonlocal term with a
\textit{more compatible} semilocal functional or by proposing a new nonlocal
term. For instance, Lee \textit{et al}. (LMKLL)\cite{LeePRB10}
proposed to modify slightly the DRSLL term (by changing the value of one
parameter) and to use it in combination with PW86R\cite{MurrayJCTC09}
(a refitted version of the Perdew-Wang functional PW86\cite{PerdewPRB86}).
Their resulting functional (called vdW-DF2) was shown to
improve over the original vdW-DF. In Ref. \onlinecite{CooperPRB10},
a new GGA exchange functional (C09$_{\text{x}}$)
was proposed to be used with the DRSLL nonlocal term. The functional,
C09$_{\text{x}}$-vdW, leads to more accurate results than vdW-DF for various
types of systems. In Refs. \onlinecite{KlimesJPCM10} and
\onlinecite{KlimesPRB11}, Klime\v{s} \textit{et al}. combined many GGA exchange
functionals (already existing or newly proposed) with the DRSLL term.
Particularly interesting are the functionals optB88-vdW and optPBE-vdW which
lead to accurate results for both finite\cite{KlimesJPCM10} and
extended\cite{KlimesPRB11} systems. We also mention
Ref. \onlinecite{WellendorffTC11}, where the nonlocal DRSLL term is used
with the GGA RPBE exchange functional\cite{HammerPRB99} and with either LDA
for correlation or a linear combination of the LDA and PBE correlation
functionals, the latter case leading to the functional named RPBEc2/3+nl
(see Table \ref{table1}).

\begin{table}
\caption{Composition of the tested exchange-correlation functionals.
The VWN5 parametrization\cite{VoskoCJP80} is used for LDA correlation.}
\label{table1}
\begin{tabular}{lccc}
\hline
\hline
Functional           & Reference                        & Exchange         & Correlation                             \\
\hline
LDA                  & \onlinecite{KohnPR65,VoskoCJP80} & LDA              & LDA                                     \\
PBE                  & \onlinecite{PerdewPRL96}         & PBE              & PBE                                     \\
vdW-DF               & \onlinecite{DionPRL04}           & revPBE           & LDA+DRSLL                               \\
vdW-DF2              & \onlinecite{LeePRB10}            & PW86R            & LDA+LMKLL                               \\
C09$_{\text{x}}$-vdW & \onlinecite{CooperPRB10}         & C09$_{\text{x}}$ & LDA+DRSLL                               \\
optB88-vdW           & \onlinecite{KlimesJPCM10}        & optB88           & LDA+DRSLL                               \\
RPBEc2/3+nl          & \onlinecite{WellendorffTC11}     & RPBE             & $\frac{1}{3}$LDA+$\frac{2}{3}$PBE+DRSLL \\
rVV10                & \onlinecite{SabatiniPRB13}       & PW86R            & PBE+rVV10                               \\
PBE-D3               & \onlinecite{GrimmeJCP10}         & PBE              & PBE+D3                                  \\
revPBE-D3            & \onlinecite{GrimmeJCP10}         & revPBE           & PBE+D3                                  \\
B97D-D3              & \onlinecite{GrimmeJCP10}         & B97D             & B97D+D3                                 \\
\hline
\hline
\end{tabular}
\end{table}

Vydrov and Van Voorhis proposed their own nonlocal functionals,
VV09\cite{VydrovPRL09} and VV10\cite{VydrovJCP10}
[also of the form given by Eq. (\ref{Ecnl})],
which were constructed such that also the short-range regime of the van der Waals
interactions is adequately described.
VV10, when added to PW86R\cite{MurrayJCTC09} exchange and
PBE correlation, has been shown to be particularly accurate
for finite systems (see, e.g., Ref. \onlinecite{HujoJCTC11}). However, its performance
for solids is rather bad,\cite{BjorkmanJPCM12,BjorkmanPRB12} therefore two
parameters of the VV10 functional were modified such that the results for solids are
improved.\cite{BjorkmanPRB12}

In this work, the results obtained for the equilibrium distance and interaction energy
of rare-gas dimers and solids will be presented. The focus will be on the performance of
several nonlocal functionals (listed in Table \ref{table1}) with which only a few
calculations on rare-gas systems have been done up to now.
The rare-gas systems are the prototypical van der Waals systems where the dispersion
interactions are the only source of attraction between atoms and for which highly accurate
\textit{ab initio} or empirical results are available. The rare-gas dimers have been used
numerous times for the testing of functionals for weak interactions
(see, e.g., Refs.
\onlinecite{PattonPRA97,PattonIJQC98,vanMourikJCP02,XuPNAS04,JohnsonCPL04,XuJCP05,TaoJCP05,ZhaoJPCA06,GerberJCP07,MurdachaewJPCA08,KannemannJCTC09,JohnsonJCP09,YousafJCTC10,RoyJCP12}
for extensive tests), while tests on rare-gas solids are less common and recent.
\cite{OrtmannPRB06,TranPRB07,HarlPRB08,TkatchenkoPRB08,HaasPRB09,MaerzkeJPCA09,BuckoJPCA10,AlSaidiJCTC12,OterodelaRozaJCP12}

\section{\label{methods}Methods}

The calculations were done with the Quickstep module\cite{VandeVondeleCPC05} of
the CP2K program package,\cite{CP2K} which is based on a mixed Gaussian and plane
waves formalism.\cite{LippertMP97} More specifically, we used the Gaussian and
augmented-plane-wave method (GAPW),\cite{LippertTCA99} which allows for all-electron
calculations. The calculations on the rare-gas dimers He$_{2}$, Ne$_{2}$,
Ar$_{2}$, and Kr$_{2}$ were done with the augmented correlation consistent
polarized quintuple zeta (aug-cc-pV5Z) basis sets,\cite{WoonJCP94,WilsonJCP99}
which lead to results very close to the basis set limit (see, e.g.,
Ref. \onlinecite{RoyJCP12}). In order to avoid the problem of linear dependence
due to diffuse functions usually experienced in solids (as in the present case),
the calculations on solid Ne, Ar, and Kr were done without the augmentation
functions by using the cc-pV5Z basis sets.\cite{WoonJCP94,WilsonJCP99}
The face-centered cubic (fcc) structure
was considered for the rare-gas solids and we checked that using a unit cell
comprising 32 atoms ($2\times2\times2$ of the fcc four-atom unit cell) gives
results which are very well converged with respect to the size of the supercell.

The nonlocal term [Eq. (\ref{Ecnl})] was implemented according to the
scheme of Rom\'{a}n-P\'{e}rez and Soler,\cite{RomanPerezPRL09} which uses
fast Fourier transforms, and therefore leads to calculations scaling as
$\mathcal{O}\left(N\log N\right)$ ($N$ is the number
of points on the grid) instead of $\mathcal{O}\left(N^{2}\right)$ for
a direct evaluation of Eq. (\ref{Ecnl}) in real space.
Note that the method of Rom\'{a}n-P\'{e}rez and Soler also leads to an efficient
calculation of the contribution of the nonlocal term to the KS-DFT potential
(needed for the forces) and stress tensor.
\cite{RomanPerezPRL09,SabatiniJPCM12}
In our implementation, the nonlocal term is evaluated using only the
smooth part of the electron density of the GAPW method. However, in Ref.
\onlinecite{KlimesPRB11}, it was shown that within the
projected-augmented wave\cite{BlochlPRB94} method, plugin the all-electron density
or the valence density into Eq. (\ref{Ecnl}) leads to very similar results.
Actually, we checked that our results agree very closely with the results
obtained with other codes when available
[Ar$_{2}$ with vdW-DF\cite{DionPRL04,ThonhauserPRB07,NabokCPC11,SabatiniPRB13} and
(r)VV10\cite{VydrovJCP10,SabatiniPRB13} and Kr$_{2}$ with vdW-DF
\cite{DionPRL04,ThonhauserPRB07,LazicCPC10} and VV10\cite{VydrovJCP10}].

In addition to the functionals already introduced in Sec. \ref{introduction},
we also mention rVV10,\cite{SabatiniPRB13} which is a revised version of
VV10\cite{VydrovJCP10} such that its evaluation can also be done with the
method of Rom\'{a}n-P\'{e}rez and Soler. It was shown (Ref.
\onlinecite{SabatiniPRB13}) that VV10 and rVV10 give very similar results.
rVV10 is among the functionals tested in the present work (Table \ref{table1}).
For comparison purposes, we also considered the standard functionals LDA and
PBE,\cite{PerdewPRL96} as well as the dispersion-corrected functionals PBE-D3,
revPBE-D3, and B97D-D3 (B97D\cite{GrimmeJCC06} is a reparametrization of
B97\cite{BeckeJCP97}), where D3 refers to the third set of parameters
$C_{n}^{AB}$ [in Eq. (\ref{Edisp})] proposed by Grimme
(the three-body term was included in our calculations).\cite{GrimmeJCP10} Note
that we used the VWN5 parametrization\cite{VoskoCJP80} for the LDA correlation.
Finally, we mention that LIBXC, a library of exchange-correlation functionals,
\cite{MarquesCPC12} has been used for the evaluation of some of the semilocal
functionals in Table \ref{table1}.

\section{\label{results}Results}

\subsection{\label{dimers}Rare-gas dimers}

\begin{table*}
\caption{Equilibrium bond length $R_{0}$ (in \AA) and interaction energy
$\Delta E$ (in meV and with opposite sign) of rare-gas dimers calculated from
various functionals and compared to accurate reference values and results
obtained from the exchange-hole dipole moment model of Becke and Johnson (BJ).}
\label{table2}
\begin{ruledtabular}
\begin{tabular}{lcccccccc}
\multicolumn{1}{}{} &
\multicolumn{2}{c}{He$_{2}$} &
\multicolumn{2}{c}{Ne$_{2}$} &
\multicolumn{2}{c}{Ar$_{2}$} &
\multicolumn{2}{c}{Kr$_{2}$} \\
\cline{2-3}\cline{4-5}\cline{6-7}\cline{8-9}
\multicolumn{1}{l}{Functional} &
\multicolumn{1}{c}{$R_{0}$} &
\multicolumn{1}{c}{$\Delta E$} &
\multicolumn{1}{c}{$R_{0}$} &
\multicolumn{1}{c}{$\Delta E$} &
\multicolumn{1}{c}{$R_{0}$} &
\multicolumn{1}{c}{$\Delta E$} &
\multicolumn{1}{c}{$R_{0}$} &
\multicolumn{1}{c}{$\Delta E$} \\
\hline
LDA                          & 2.40 &  9.6 & 2.64 & 20.4 & 3.40 & 30.9 & 3.68 & 36.7 \\
PBE                          & 2.76 &  3.2 & 3.08 &  5.6 & 4.00 &  6.3 & 4.36 &  6.9 \\
vdW-DF                       & 2.82 &  6.6 & 3.07 & 14.1 & 3.92 & 23.1 & 4.27 & 26.2 \\
vdW-DF2                      & 2.75 &  2.8 & 2.95 &  9.2 & 3.75 & 18.3 & 4.09 & 22.3 \\
C09$_{\text{x}}$-vdW         & 3.19 &  4.1 & 3.51 &  6.6 & 4.37 & 11.5 & 4.71 & 13.4 \\
optB88-vdW                   & 3.48 &  0.5 & 3.30 &  3.0 & 3.93 & 11.7 & 4.20 & 16.1 \\
RPBEc2/3+nl                  & 2.66 & 11.2 & 2.93 & 23.1 & 3.77 & 34.4 & 4.10 & 38.0 \\
rVV10                        & 2.92 &  0.9 & 3.01 &  5.6 & 3.73 & 13.9 & 4.00 & 19.3 \\
PBE-D3                       & 2.66 &  5.7 & 3.01 &  9.9 & 3.88 & 15.3 & 4.16 & 19.3 \\
revPBE-D3                    & 2.90 &  3.0 & 3.20 &  5.6 & 3.93 & 12.8 & 4.18 & 17.9 \\
B97D-D3                      & 3.01 &  2.4 & 3.33 &  4.3 & 3.99 & 11.3 & 4.18 & 17.2 \\
PW86xPBEc-BJ\footnotemark[1] & 3.01 &  0.8 & 3.12 &  3.8 & 3.84 & 11.2 & 4.07 & 17.0 \\ 
Reference\footnotemark[2]    & 2.97 &  0.9 & 3.09 &  3.6 & 3.76 & 12.4 & 4.01 & 17.4 \\
\end{tabular}
\footnotetext[1]{Reference \onlinecite{KannemannJCTC09}.}
\footnotetext[2]{Reference \onlinecite{TangJCP03}.}
\end{ruledtabular}
\end{table*}

\begin{figure*}
\begin{picture}(16,16)(0,0)
\put(0,8){\epsfxsize=7cm \epsfbox{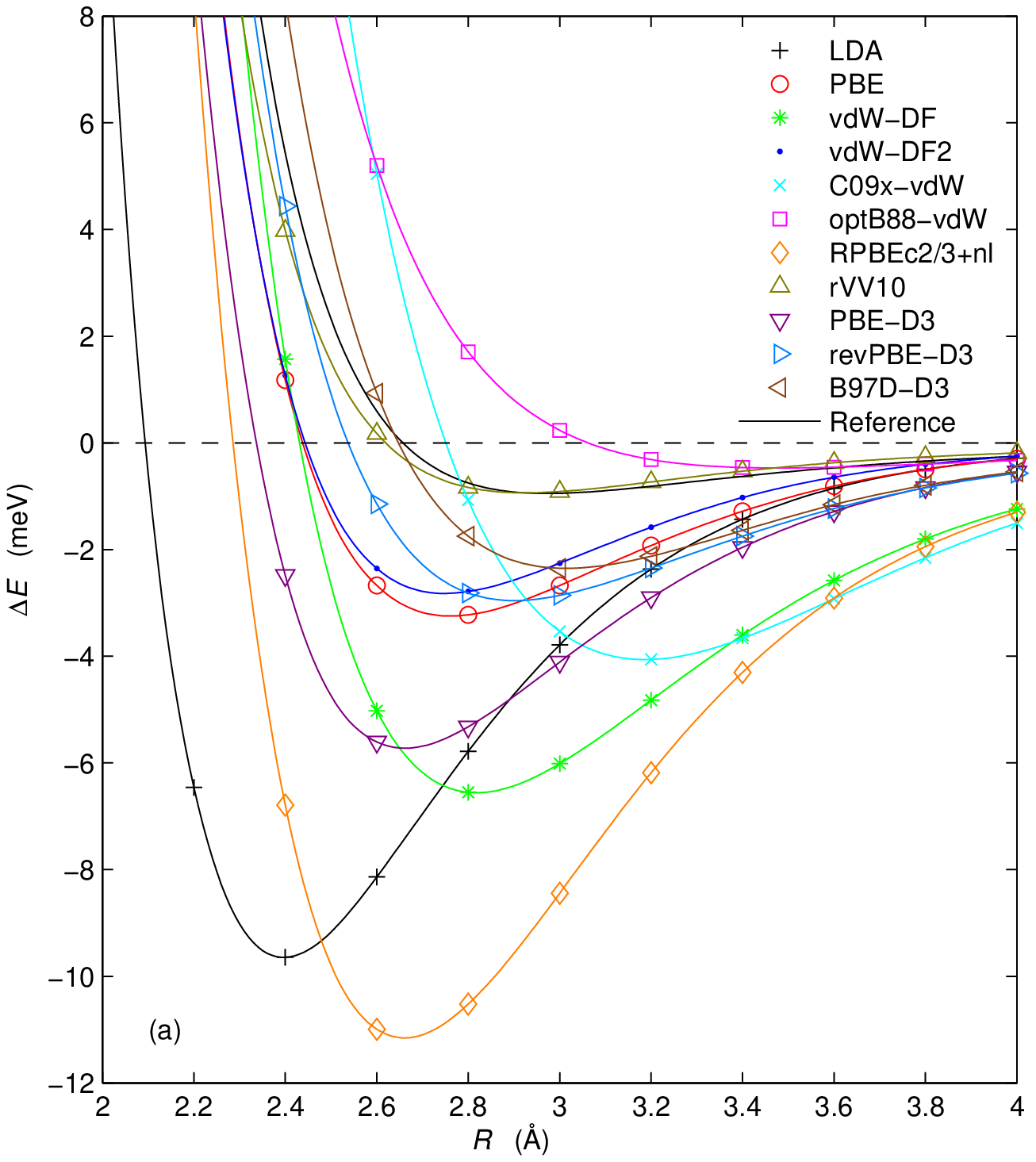}}
\put(8,8){\epsfxsize=7cm \epsfbox{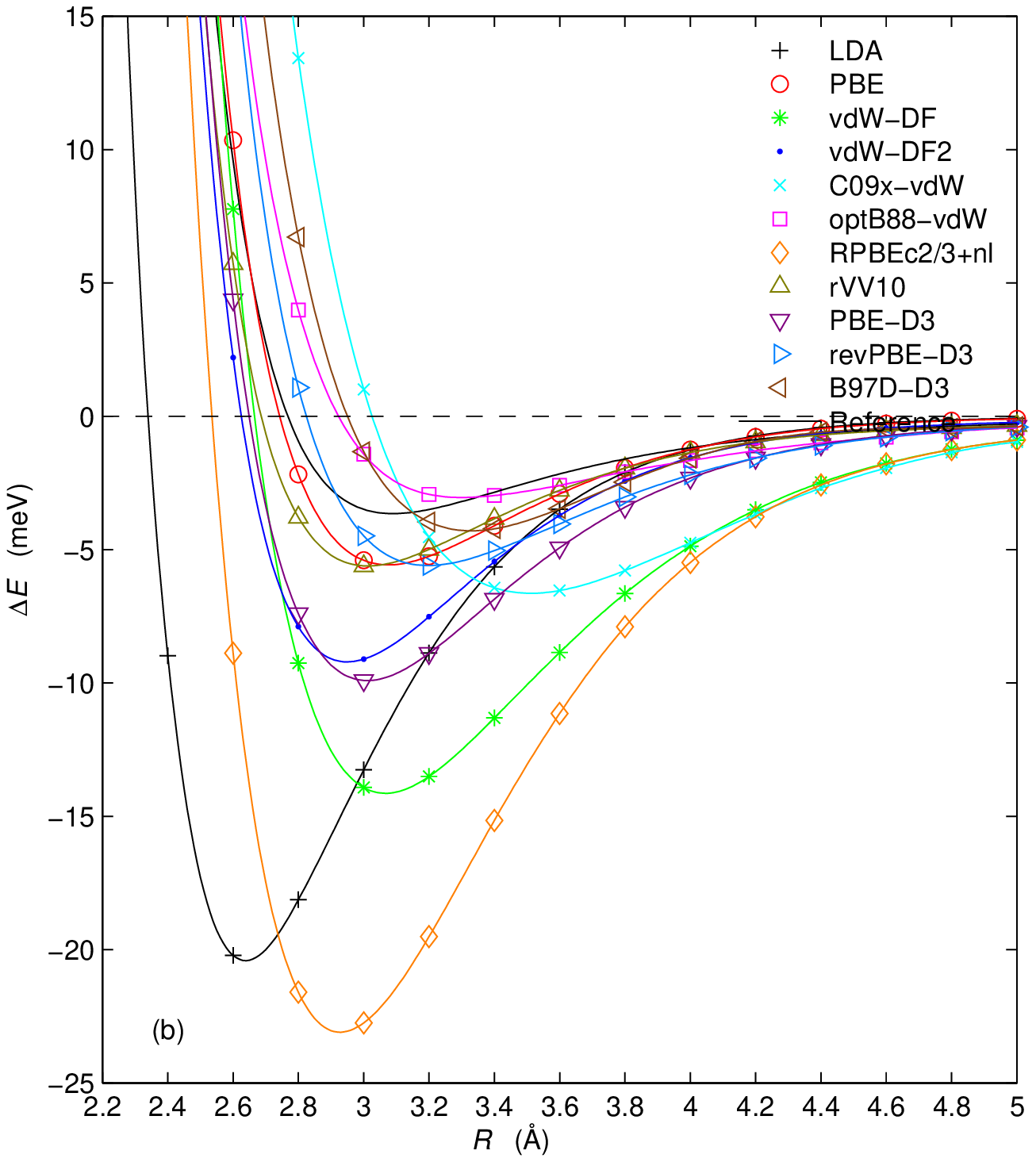}}
\put(0,0){\epsfxsize=7cm \epsfbox{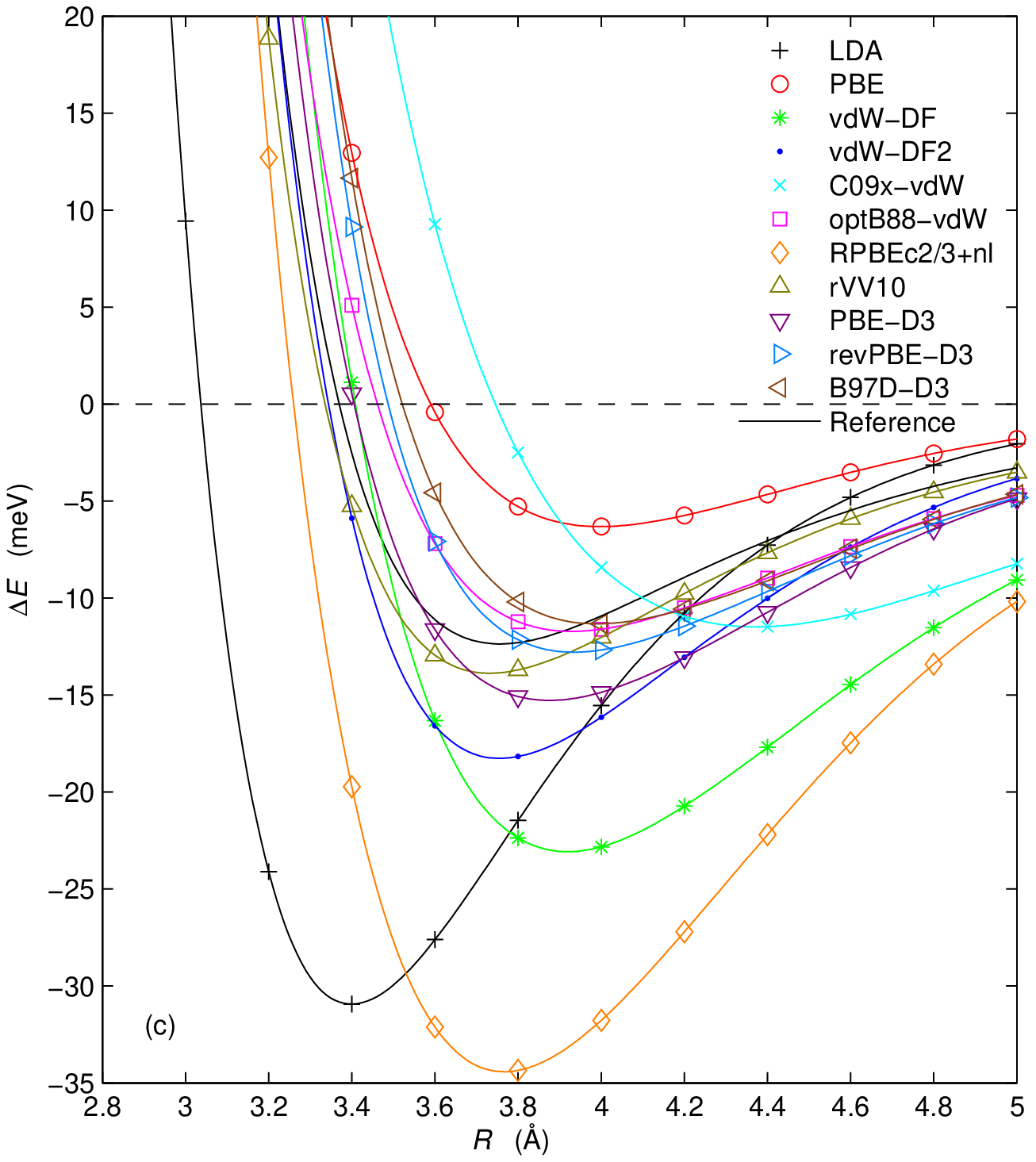}}
\put(8,0){\epsfxsize=7cm \epsfbox{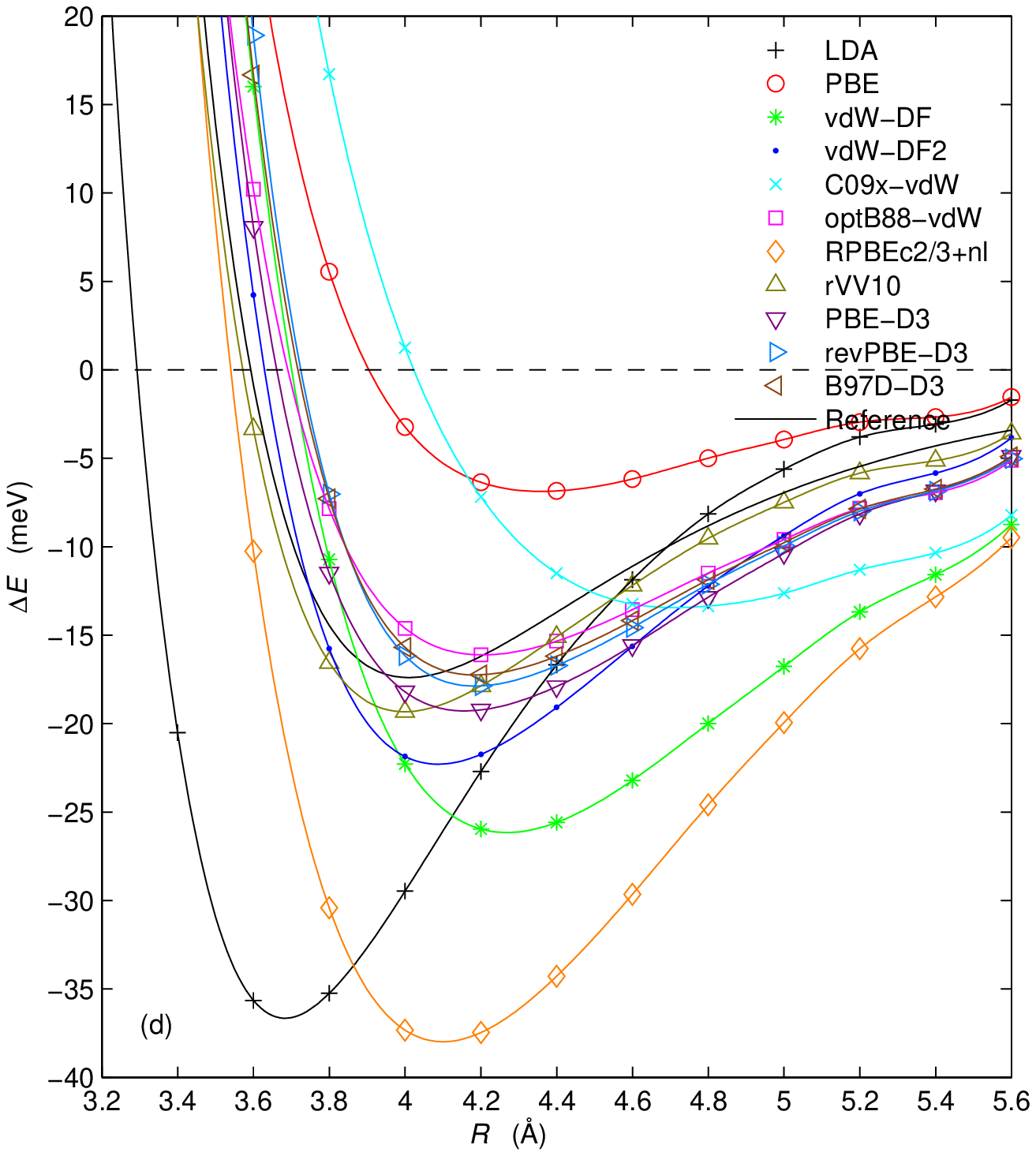}}
\end{picture}
\caption{\label{fig1}Interaction energy curves for
(a) He$_{2}$, (b) Ne$_{2}$, (c) Ar$_{2}$, and (d) Kr$_{2}$ obtained from
various functionals and compared to reference results\cite{TangJCP03}
(black line without symbols).}
\end{figure*}

The interaction energy curves of the rare-gas dimers He$_{2}$, Ne$_{2}$,
Ar$_{2}$, and Kr$_{2}$ are displayed in Fig. \ref{fig1} and the corresponding
values at the minimum (equilibrium distance $R_{0}$ and binding energy
$\Delta E$) are shown in Table \ref{table1}. The KS-DFT results are compared
to very accurate reference (theoretical or experimental,
see Ref. \onlinecite{TangJCP03} for details) results.

Discussing first the results obtained with the (semi)local functionals,
it is already known\cite{KristyanCPL94,PerezJordaCPL95} that LDA strongly
underestimates the bond lengths and
overestimates the binding energies of all dimers, this trend being systematically
observed with LDA for intermolecular complexes. For He$_{2}$ and Ne$_{2}$, LDA
leads to binding energies which are one order of magnitude too large and to
distances which are about 0.5 \AA~too small.
Among the countless semilocal and hybrid functionals tested on rare-gas dimers,
PBE (a GGA free of any empirical parameter) is one of the most accurate
(or least inaccurate, see Refs. \onlinecite{JohnsonCPL04,XuJCP05,ZhaoJPCA06}
for extensive tests). Still, PBE accuracy can not be considered as satisfying,
Ne$_{2}$ excepted, since it largely overbinds He$_{2}$ and underbinds Ar$_{2}$ and
Kr$_{2}$. In the group of semilocal and hybrid functionals (which do not
include the physics of dispersion interactions), it was shown\cite{ZhaoJPCA06}
that the hybrid B97-1\cite{HamprechtJCP98} and meta-GGA hybrid
M05-2X\cite{ZhaoJCTC06} are also among the best for rare-gas dimers, but as in
the case of PBE the results are in some cases rather inaccurate.
Finally, from all previous studies on rare-gas dimers, we can conclude that
there is no semilocal or hybrid functional that can be considered as reliable.

Turning now to the results obtained with the six nonlocal functionals, we can see
that a large range of results can be obtained. vdW-DF largely overbinds all four
dimers, and while the bond length is reasonable for He$_{2}$ and Ne$_{2}$,
it is too large by 0.2$-$0.3 \AA~for Ar$_{2}$ and Kr$_{2}$. vdW-DF2 improves over
vdW-DF for $\Delta E$ by reducing the overbinding by a factor of two, but now
the bond lengths of He$_{2}$ and Ne$_{2}$ are clearly too short.
The bond lengths obtained with the C09$_{\text{x}}$-vdW functional
are as inaccurate as the LDA ones, but with the opposite trend (overestimation
ranging from $0.2$ \AA~for He$_{2}$ to $0.7$ \AA~for Kr$_{2}$).
The C09$_{\text{x}}$-vdW interaction energies are not particularly accurate
except for Ar$_{2}$. optB88-vdW leads to quite accurate results for the binding
energy $\Delta E$, however, the bond lengths $R_{0}$ are too large
(in particular for He$_{2}$ with 0.5 \AA~of error). The binding energies obtained
with the RPBEc2/3+nl functional constitute a disaster since they are even
larger than LDA values, while the equilibrium bond length is accurate
for Ar$_{2}$ and Kr$_{2}$, but not for the two lighter dimers. Among all
tested functionals in this work, rVV10 is clearly the most accurate one.
From Fig. \ref{fig1}(a), we can see that for He$_{2}$ the rVV10 and reference
curves coincide very closely along the whole range of considered intermolecular
distances and correspond to the same binding energy (0.9 meV). For the other
dimers, also both the bond lengths and interaction energies are very accurate.
The largest error in $\Delta E$ is only 2 meV (for Ne$_{2}$ and Kr$_{2}$).

Concerning the three DFT-D3 methods that we considered, revPBE-D3 leads to a
rather accurate bond length for He$_{2}$, but overestimates $\Delta E$. For
the other three dimers, revPBE-D3 yields values for $R_{0}$, which are too large
by 0.1$-$0.2 \AA, but quite accurate values for $\Delta E$. Overall, PBE-D3
leads to values which are less satisfying than revPBE-D3.
B97D-D3 leads to results which are quite similar to revPBE-D3, but overestimates
the bond lengths even more for Ne$_{2}$ and Ar$_{2}$.
When compared to the nonlocal functionals, revPBE-D3 and B97D-D3 seem to show
more stability in the results, except when compared to rVV10 which is by far
the most accurate functional.

Among the previously published works on the testing of DFT functionals
on rare-gas dimers, we should mention the results obtained by Kannemann and
Becke\cite{KannemannJCTC09} with the functional PW86xPBEc-BJ (results also
shown in Table \ref{table2}), where BJ refers to the XDM model for dispersion
of Becke and Johnson.\cite{BeckeJCP05} Their calculated bond lengths
and binding energies are in very close agreement with the reference results,
and actually the accuracy of rVV10 and PW86xPBEc-BJ can be considered as
similar. However, it is important to note that the two adjustable parameters
in the PW86xPBEc-BJ functional were determined by minimizing the error for
$\Delta E$ of a set of ten rare-gas dimers (all combinations involving
He, Ne, Ar, and Kr). rVV10 (VV10) also contains two parameters,
but one of them was adjusted such that the mean error of $C_{6}^{AA}$
coefficients for a set of 54 species (among them He, Ne, Ar, and Kr) is
minimized,\cite{VydrovJCP10}
while the other was determined using the S22 set of noncovalent complexes,
\cite{JureckaPCCP06} which does not contain any rare-gas atoms.
Therefore, (r)VV10 was certainly not adjusted exclusively on rare-gas systems,
which makes its excellent performances on these systems even more impressive.

\subsection{\label{solids}Rare-gas solids}

\begin{table*}
\caption{Equilibrium lattice constant $a_{0}$ (in \AA) and cohesive energy
$\Delta E$ (in meV/atom and with opposite sign) of rare-gas solids calculated
from various functionals and compared to reference [CCSD(T)] results as well as
the PBE-TS and RPA methods.}
\label{table3}
\begin{ruledtabular}
\begin{tabular}{lcccccc}
\multicolumn{1}{}{} &
\multicolumn{2}{c}{Ne} &
\multicolumn{2}{c}{Ar} &
\multicolumn{2}{c}{Kr} \\
\cline{2-3}\cline{4-5}\cline{6-7}
\multicolumn{1}{l}{Functional} &
\multicolumn{1}{c}{$a_{0}$} &
\multicolumn{1}{c}{$\Delta E$} &
\multicolumn{1}{c}{$a_{0}$} &
\multicolumn{1}{c}{$\Delta E$} &
\multicolumn{1}{c}{$a_{0}$} &
\multicolumn{1}{c}{$\Delta E$} \\
\hline
LDA                      & 3.86  &  92 & 4.94  & 136 & 5.36  & 164 \\
PBE                      & 4.55  &  25 & 5.93  &  25 & 6.42  &  25 \\
vdW-DF                   & 4.32  & 101 & 5.49  & 163 & 5.96  & 184 \\
vdW-DF2                  & 4.17  &  65 & 5.29  & 130 & 5.75  & 157 \\
C09$_{\text{x}}$-vdW     & 4.90  &  51 & 6.00  &  83 & 6.39  &  96 \\
optB88-vdW               & 4.24  &  59 & 5.24  & 143 & 5.63  & 181 \\
RPBEc2/3+nl              & 4.19  & 146 & 5.35  & 222 & 5.80  & 246 \\
rVV10                    & 4.19  &  49 & 5.17  & 117 & 5.53  & 162 \\
PBE-D3                   & 4.37  &  53 & 5.58  &  84 & 5.93  & 108 \\
revPBE-D3                & 4.66  &  32 & 5.62  &  71 & 5.89  & 104 \\
B97D-D3                  & 4.78  &  26 & 5.69  &  66 & 5.87  & 104 \\
PBE-TS\footnotemark[1]   & 4.42  &  43 & 5.51  &  83 & 5.90  &  97 \\
RPA(PBE)\footnotemark[2] & 4.5   &  17 & 5.3   &  83 & 5.7   & 112 \\
CCSD(T)\footnotemark[3]  & 4.297 &  26 & 5.251 &  88 & 5.598 & 122 \\
\end{tabular}
\footnotetext[1]{Reference \onlinecite{AlSaidiJCTC12}.}
\footnotetext[2]{RPA energy evaluated with PBE orbitals and
eigenvalues.\cite{HarlPRB08}}
\footnotetext[3]{Reference \onlinecite{RosciszewskiPRB00}.}
\end{ruledtabular}
\end{table*}

\begin{figure*}
\begin{picture}(17.4,8)(0,0)
\put(0,0){\epsfxsize=5.8cm \epsfbox{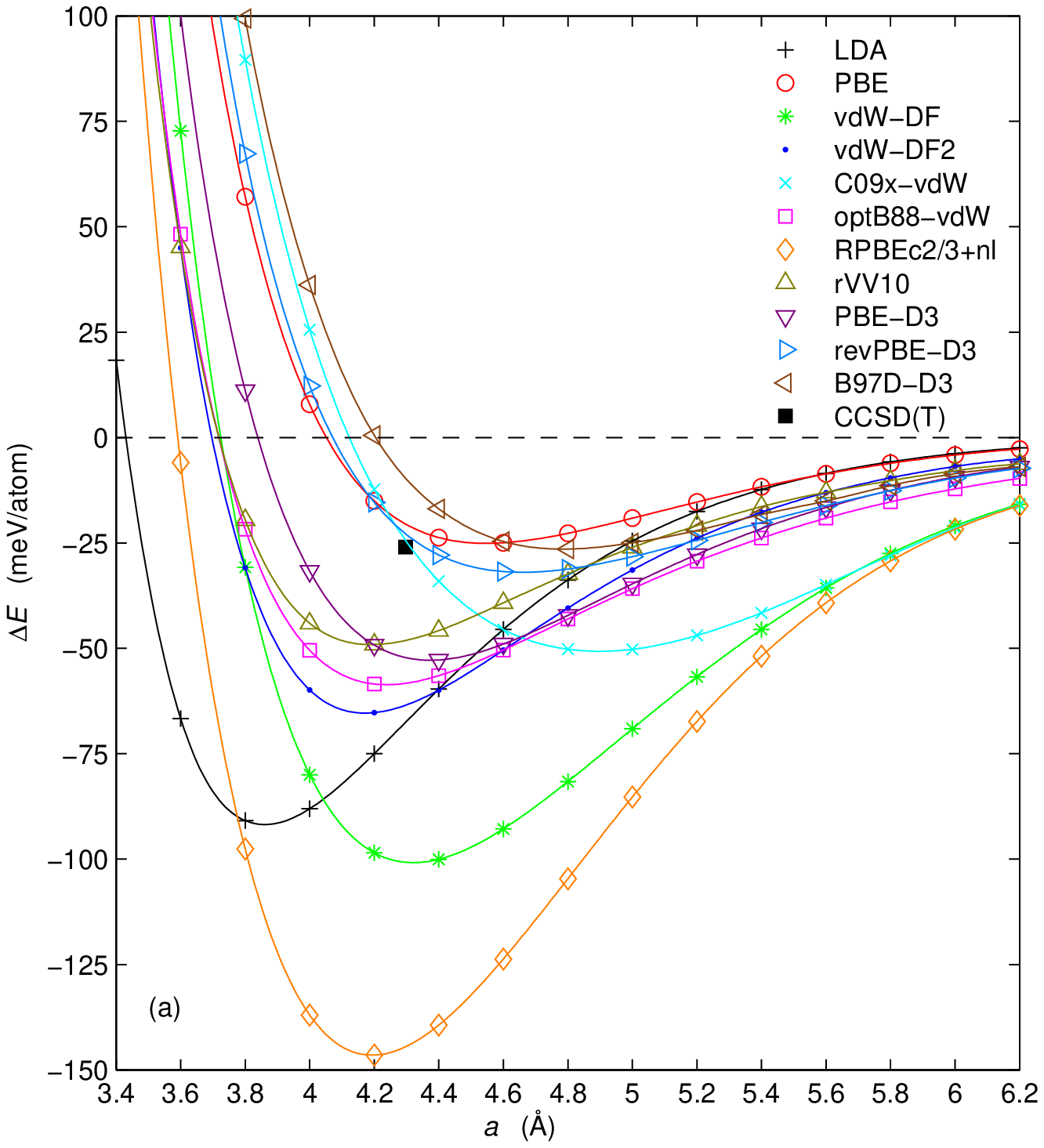}}
\put(5.8,0){\epsfxsize5.8cm \epsfbox{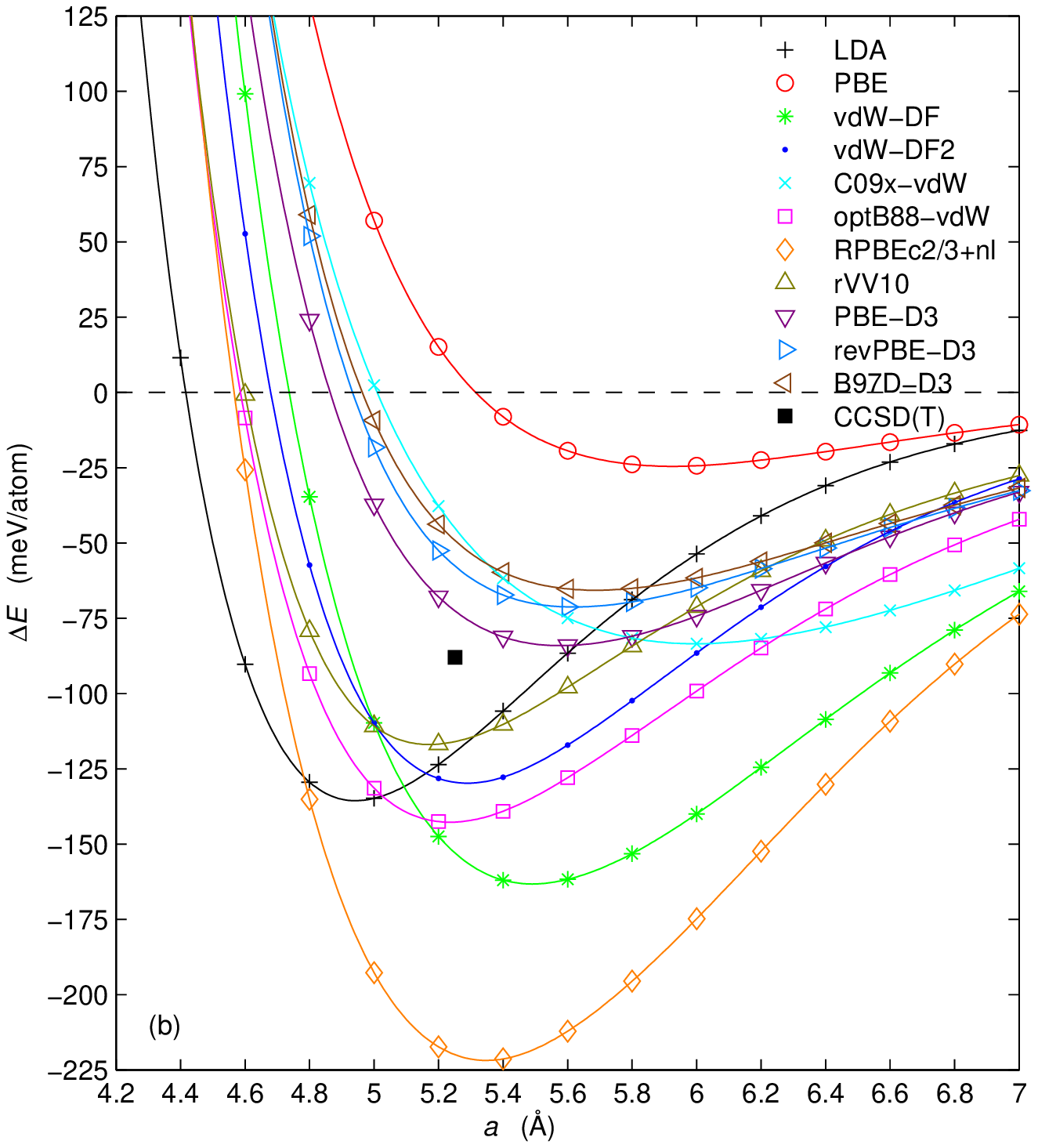}}
\put(11.6,0){\epsfxsize=5.8cm \epsfbox{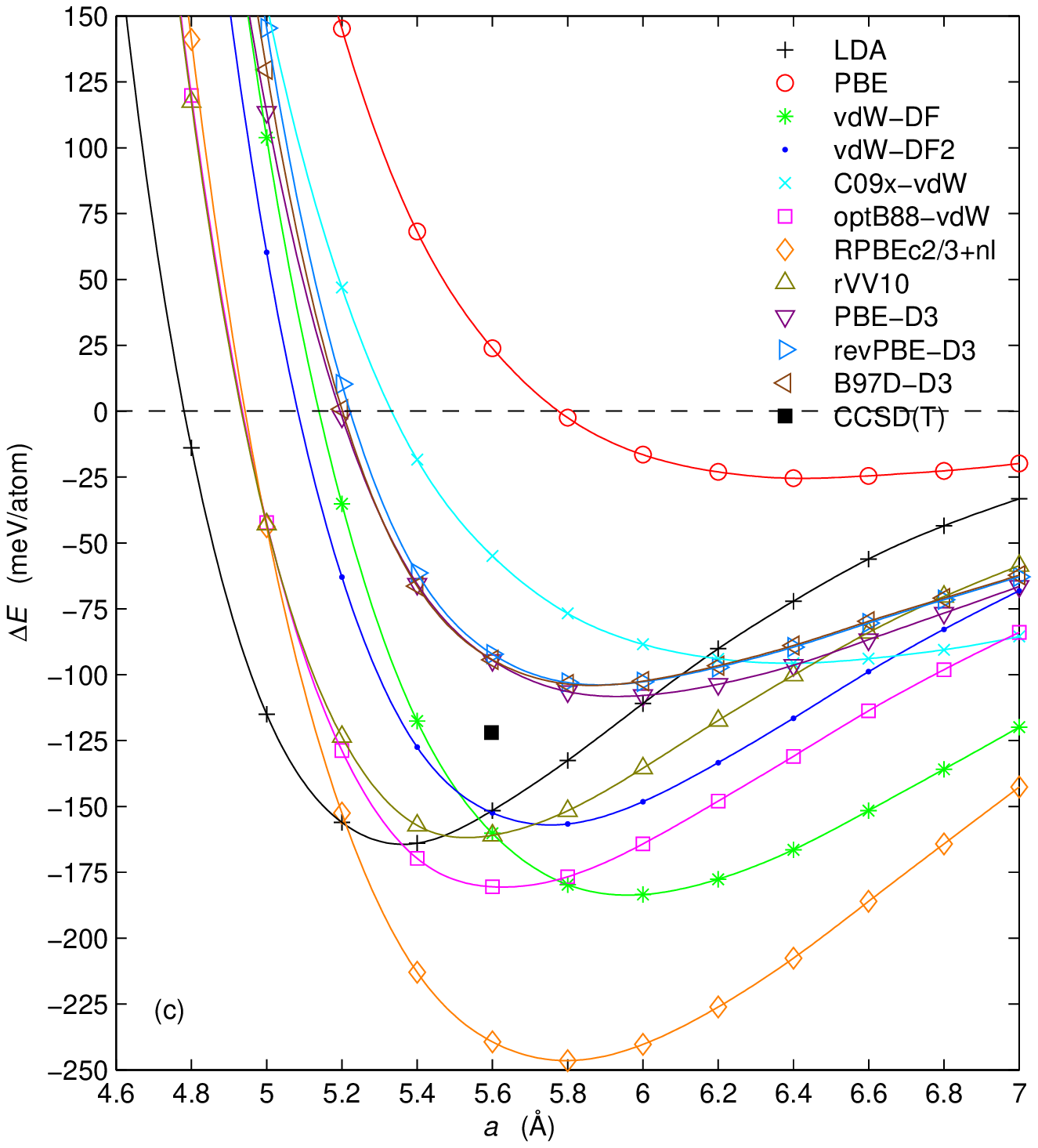}}
\end{picture}
\caption{\label{fig2}
Cohesive energy curves for (a) Ne, (b) Ar, and (c) Kr obtained from
various functionals and compared to CCSD(T) results\cite{RosciszewskiPRB00}
(black square).}
\end{figure*}

The results for the rare-gas solids Ne, Ar, and Kr are shown in Fig. \ref{fig2}
and Table \ref{table3}. The reference results were obtained from coupled
cluster with single, double, and perturbative triple excitations [CCSD(T)]
calculations.\cite{RosciszewskiPRB00} For a meaningful comparison of our KS-DFT
results with the CCSD(T) results, the zero-point energy calculated in
Ref. \onlinecite{RosciszewskiPRB00} has been removed from the CCSD(T) results.

As for the dimers, LDA leads to severe underestimation of the lattice constant
$a_{0}$ and overestimation of the cohesive energy $\Delta E$ for all three
solids. Note, however, that in some cases LDA can, at a qualitative level,
give relatively correct results
for solids which are bound by weak interactions. Such examples include layered
solids like graphite (see, e.g., Refs. \onlinecite{HasegawaPRB04} and
\onlinecite{TranPRB07}). As already observed in Ref. \onlinecite{HarlPRB08},
PBE gives essentially the same cohesive energy (25 meV/atom) for the
three solids (very large underestimation for Ar and Kr).
This is somewhat similar to what is observed for the
corresponding dimers (see Table \ref{table2}). The PBE lattice constants
are by far too large (by more than 0.7 \AA~for Ar and Kr).

Concerning the nonlocal functionals based on the DRSLL or LMKLL kernels, the
observations are the following. vdW-DF and vdW-DF2 clearly overbind the
rare-gas solids, C09$_{\text{x}}$-vdW totally fails for the lattice constant,
and RPBEc2/3+nl leads to the largest overbinding (as for the dimers).
optB88-vdW leads to quite accurate values for $a_{0}$, but
overestimates the cohesive energy rather strongly (50\%$-$100\%), while in the
case of the dimers optB88-vdW was quite good for the interaction energy.

The rVV10 nonlocal functional seems to be again superior to the other nonlocal
functionals. The bond lengths are rather good (albeit too short by
$\sim0.1$ \AA~for Ne), while the cohesive energies are too large for all
three solids, but the error is smaller than for the other nonlocal functionals
except C09$_{\text{x}}$-vdW.
The cohesive energies obtained with the DFT-D3 methods are quite accurate
(PBE-D3 for Ne excepted), however the lattice constants are consistently too
large by more than 0.3 \AA~in most cases.

Also shown in Table \ref{table2} are the results from
Ref. \onlinecite{AlSaidiJCTC12} obtained with the Tkatchenko and
Scheffler\cite{TkatchenkoPRL09} (TS) approach using PBE for the semilocal part.
We can see that the PBE-TS results are similar to the results from PBE-D3
for both the lattice constant and cohesive energy. For completeness, we also show in
Table \ref{table2} the values from the \textit{non-DFT} method RPA
(random-phase approximation).\cite{HarlPRB08} The RPA bond lengths are somehow
overestimated, but the cohesive energies are very close to the CCSD(T) values.
The RPA method is superior to the KS-DFT methods considered in the present
work, but leads to calculations which are obviously much more expensive.
Finally, we mention the DFT+XDM results from
Ref. \onlinecite{OterodelaRozaJCP12} for the lattice constant $a_{0}$, where
the XDM dispersion correction was added to two different GGA functionals.
The results are good only for Kr, whereas for Ne and Ar rather inaccurate
values were obtained.

As a summary of the results on rare-gas solids, the rVV10 nonlocal functional
seems to be a relatively good choice (at least compared to the other
functionals), but leads to non-negligible overestimations of the cohesive
energy, and this more than in the case of the rare-gas dimers.

\section{\label{discussion}Further Discussion}

\subsection{\label{decomposition}Energy decomposition}

\begin{figure}
\includegraphics[scale=0.9]{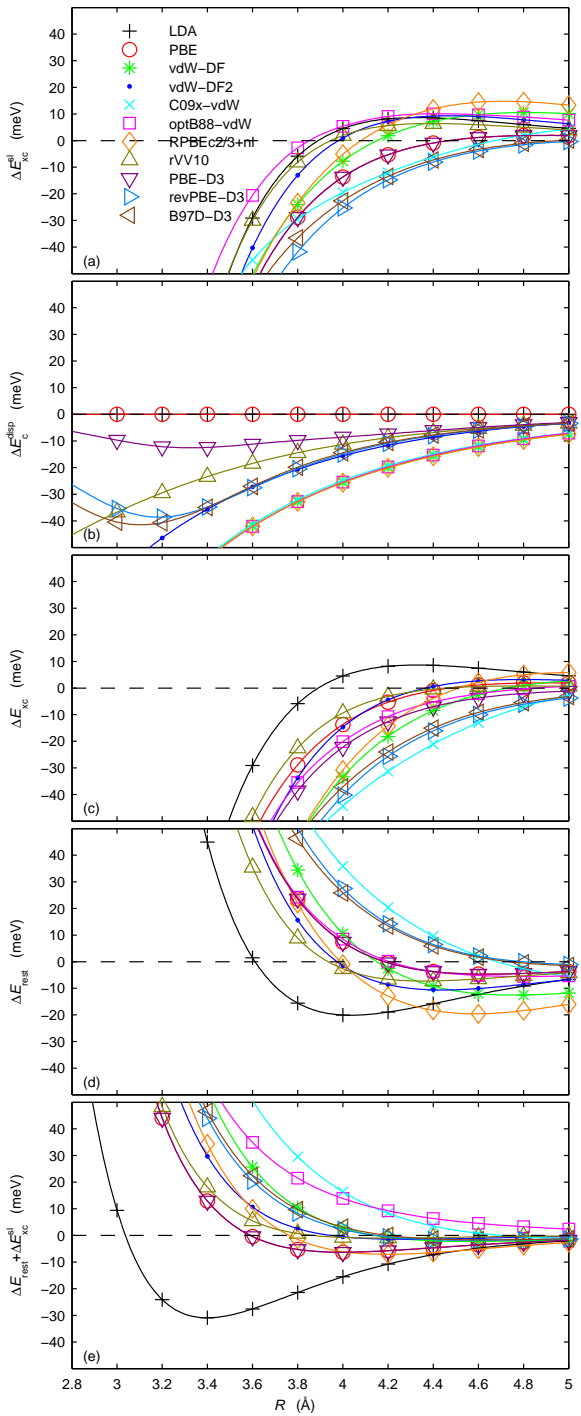}
\caption{\label{fig3}
Contributions to the Ar$_{2}$ interaction energy coming from
(a) the semilocal (sl) xc functional, (b) the dispersion energy
[Eqs. (\ref{Edisp}) or (\ref{Ecnl})], (c) the total (sl plus dispersion) xc
functional, (d) the rest (kinetic plus electrostatic), and (e) the sum
of sl and rest. The addition of (c) and (d) gives the total interaction energy
of Fig. \ref{fig1}(c).}
\end{figure}

\begin{figure}
\includegraphics[scale=0.9]{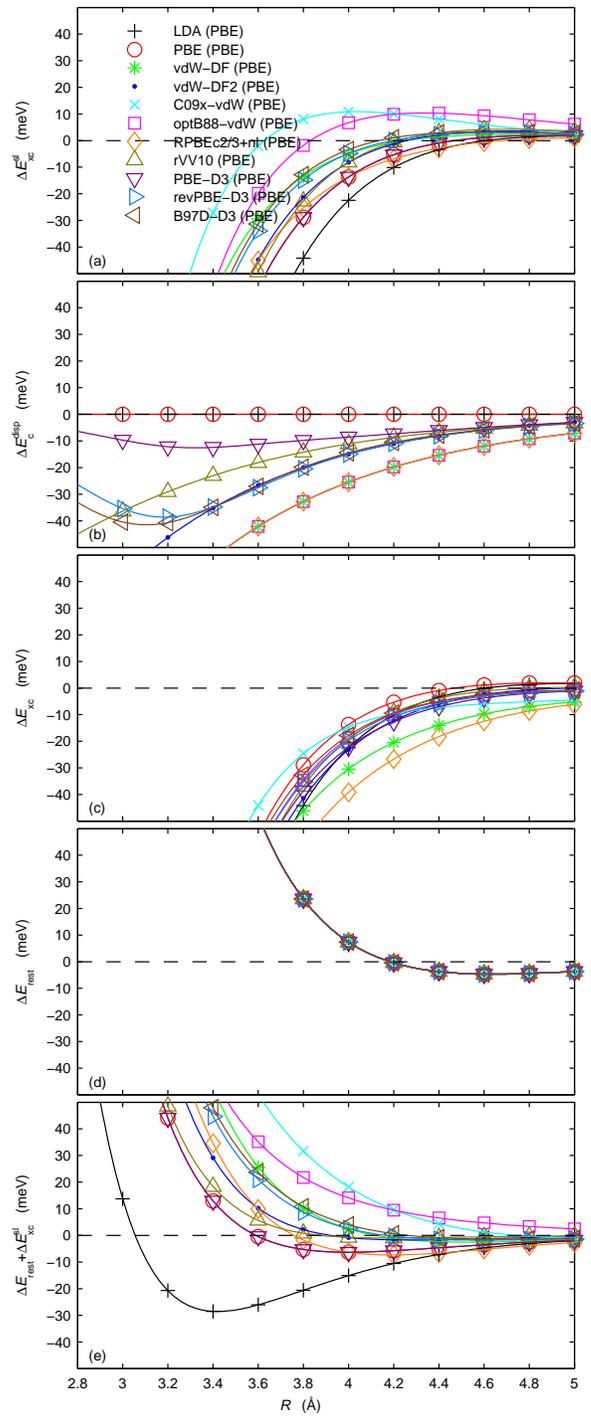}
\caption{\label{fig4}
Same as Fig. \ref{fig3}, but obtained from non-self-consistent calculations by
plugin the PBE orbitals and electron density into the functionals.}
\end{figure}

In order to have more insight into the results of Secs. \ref{dimers} and
\ref{solids}, we now consider the various contributions to the interaction
energy $\Delta E$ and their relative importance. Figure \ref{fig3} shows for
the Ar dimer (similar trends are observed for the other rare-gas
dimers and solids) the contributions to $\Delta E$ coming from the
(semi)local exchange-correlation functional
($\Delta E_{\text{xc}}^{\text{sl}}$), the (atom-pairwise or nonlocal) dispersion
energy term ($\Delta E_{\text{c}}^{\text{disp}}$), the sum of these two
($\Delta E_{\text{xc}}=\Delta E_{\text{xc}}^{\text{sl}}+\Delta E_{\text{c}}^{\text{disp}}$),
the rest of the terms ($\Delta E_{\text{rest}}$) representing the kinetic
and electrostatic energies, and the sum
$\Delta E_{\text{rest}}+\Delta E_{\text{xc}}^{\text{sl}}$.
The sum of $\Delta E_{\text{xc}}$ and
$\Delta E_{\text{rest}}$ gives the total interaction energy $\Delta E$ shown
in Fig. \ref{fig1}(c).

Around the experimental equilibrium bond length ($\sim3.8$ \AA) all terms seem
to be of roughly equal importance and the absolute values range from 10 to 50
meV depending on the functional. However, for smaller bond lengths $R$, the
curves $\Delta E_{\text{xc}}^{\text{sl}}$ and $\Delta E_{\text{rest}}$ vary faster than
$\Delta E_{\text{c}}^{\text{disp}}$ and these terms become much more important.
For instance, at $R=2.0$ \AA~(not shown), the magnitude of $\Delta E_{\text{rest}}$
and $\Delta E_{\text{xc}}^{\text{sl}}$ is around 8000 and 4000 meV, respectively, while
for $\Delta E_{\text{c}}^{\text{disp}}$ it is smaller than 5 meV for the DFT-D3
methods and between 100 and 300 meV for the nonlocal methods. Note that the
different behavior of $\Delta E_{\text{c}}^{\text{disp}}$ for the DFT-D3 method at small
values of $R$ is due to the damping function $f_{n}^{\text{damp}}$ in
Eq. (\ref{Edisp}).

In Fig. \ref{fig3}(b), we can see that among the nonlocal functionals, rVV10
and DRSLL lead to the smallest and largest values (in magnitude) for
$\Delta E_{\text{c}}^{\text{disp}}$, respectively. In the range of intermolecular distances
that we considered,
$\Delta E_{\text{c}}^{\text{disp}}$ is negative, however, the energy
(the component of the total energy) is always
positive in the case of the nonlocal functionals [Eq. (\ref{Ecnl})], while the
values are negative for the atom-pairwise DFT-D3 method [Eq. (\ref{Edisp})].

The sum of the terms $\Delta E_{\text{xc}}^{\text{sl}}$ and
$\Delta E_{\text{rest}}$, which represents the interaction energy of Ar$_{2}$
calculated without the dispersion term, is shown in Fig. \ref{fig3}(e). Among
the semilocal functionals only PBE and RPBEc2/3 yield reasonable interaction
energies, while all other functionals, except LDA barely bind or do not bind at
all the two Ar atoms. Actually, it is clear that in order to avoid overbinding
(due to double counting) when using a dispersion term in the total-energy
expression, it should be combined with a semilocal functional which leads to
strongly underestimated interaction energy.\cite{DionPRL04}

From Fig. \ref{fig3}, we can also infer that the differences in $\Delta E$
between the functionals can not be understood by looking exclusively
at the contribution from the exchange-correlation energy. Indeed, the curves
for $\Delta E_{\text{rest}}$ (whose analytical form is the same for all
functionals) show differences which are as strong as for
$\Delta E_{\text{xc}}^{\text{sl}}$ and $\Delta E_{\text{c}}^{\text{disp}}$.
Actually, the differences in the $\Delta E_{\text{rest}}$ curves are a
reflection of the corresponding exchange-correlation potentials
$v_{\text{xc}}=\delta E_{\text{xc}}/\delta\rho$ used in the
KS-DFT equations. It is known that for the calculation of
properties depending on total energies, the results usually
do not depend too sensitively on the orbitals and electron density plugged
into the total-energy functional (but one needs to be very careful with this statement).
However, the individual components of the total energy show much stronger
sensitivity, but these variations tend to cancel among the different terms.
For Ar$_{2}$, we also performed non-self-consistent calculations by evaluating
all functionals with the PBE orbitals and electron density (results shown in Fig. \ref{fig4}).
The resulting equilibrium bond lengths and binding energies are essentially
the same as their self-consistent counterparts. However, from
Fig. \ref{fig4}(a) we can see that for some functionals
(LDA and C09$_{\text{x}}$-vdW in particular), the $\Delta E_{\text{xc}}^{\text{sl}}$
curve is quite different to the one obtained self-consistently
[Fig. \ref{fig3}(a)]. For the nonlocal dispersion terms (DRSLL, LMKLL, and
rVV10), basically no difference between the self-consistent and
non-self-consistent calculations can be seen, which is maybe due to the fact
that these terms are evaluated only with the smooth part of the electron density.

\subsection{\label{Threebody}Three-body interaction energy}

\begin{figure}
\includegraphics[scale=0.6]{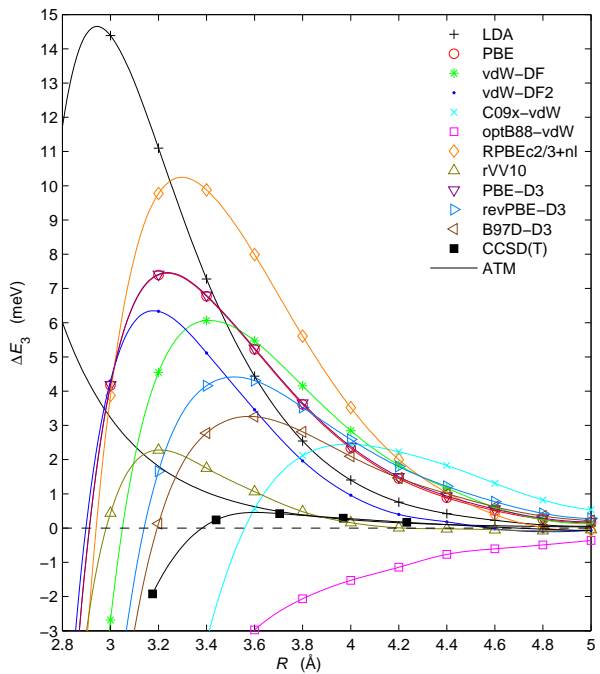}
\caption{\label{fig5}Three-body interaction energy of
Ar$_{3}$ (equilateral triangular configuration) plotted against the Ar-Ar
distance. The CCSD(T) results are from Ref. \onlinecite{PodeszwaJCP07}.
ATM (black line without symbols) is the Axilrod-Teller-Muto term given by Eq. (\ref{E3ATM}).}
\end{figure}

The leading term in the many-body contribution to the interaction energy is the
three-body nonadditive energy $\Delta E_{3}$. In the case of a simple trimer this term
is calculated as the atomization energy of the trimer minus the sum of the
atomization energies of the three dimers. If the three atoms in the
trimer are identical, then
\begin{equation}
\Delta E_{3} = E_{\text{tot}}^{\text{trimer}} - 3E_{\text{tot}}^{\text{dimer}} +
3E_{\text{tot}}^{\text{atom}}.
\label{DE3}
\end{equation}
The asymptotic behavior of the dispersion component of $\Delta E_{3}$ is given
by the Axilrod-Teller-Muto\cite{AxilrodJCP43,MutoPPMSJ43} (ATM) triple-dipole term
\begin{equation}
\Delta E_{3}^{\text{ATM}} = C_{9}^{ABC}
\frac{1+\cos(\theta_{ABC})\cos(\theta_{BCA})\cos(\theta_{CAB})}
{(R_{AB}R_{BC}R_{CA})^3},
\label{E3ATM}
\end{equation}
where $\theta_{ijk}$ and $R_{ij}$ are the angles and side lengths of the
triangle formed by the trimer and $C_{9}^{ABC}$ is the triple-dipole constant.

Using Eq. (\ref{DE3}), we calculated $\Delta E_{3}$ for the Ar trimer at
equilateral geometry, and in Fig. \ref{fig5} the results are compared to the
accurate CCSD(T) values from Ref. \onlinecite{PodeszwaJCP07} as well as the
asymptotic ATM term [Eq. (\ref{E3ATM})] with $C_{9}^{ABC}=521.7$ au.\cite{ThakkarJCP92}
In general, the three-body interaction energy $\Delta E_{3}$ of a trimer in this
configuration is the largest contribution to the many-body cohesive energy of
the corresponding solid in the fcc structure.
For interatomic distances $R$ larger than $\sim3.5$ \AA,
we can see that most functionals strongly overestimate
(too positive values) the three-body energy $\Delta E_{3}$. The exceptions
are optB88-vdW which leads to negative values for the whole range of
interatomic distances $R$ that we considered and rVV10 which seems to be
the best of the considered functionals. Closely around the equilibrium
interatomic distances in the dimer and solid ($\sim3.7$$-$$3.75$ \AA),
the rVV10 values are close to the CCSD(T) and ATM values.
However, the maximum of the $\Delta E_{3}$ curve is at $\sim3.2$ \AA~for rVV10
(and overestimated), while it is at $\sim3.7$ \AA~for CCSD(T).
In Refs. \onlinecite{TkatchenkoPRB08} and \onlinecite{MaerzkeJPCA09},
other functionals were considered for the calculation of $\Delta E_{3}$, but
none of them lead to results in qualitative agreement with the CCSD(T) results.

\section{\label{summary}Summary}

We have presented the results of KS-DFT calculations on rare-gas dimers and
solids. The focus was on the performance of nonlocal vdW functionals for the
equilibrium bond length and binding energy. The main conclusions are
(a) overall the rVV10 functional is the one performing the best, (b) some
others (e.g., C09$_{\text{x}}$-vdW or RPBEc2/3+nl) can perform very badly,
and (c) the considered DFT-D3 methods show good accuracy for the interaction
energy, but seem to lead to some (slight) overestimation of the bond lengths.
In Sec. \ref{decomposition}, we presented an analysis by decomposing the
interaction energy into its components in order to estimate the relative
importance of each term, and in Sec. \ref{Threebody} it was shown that rVV10
gives also reasonable values (at least close to the equilibrium geometry)
for the three-body nonadditive energy, while the other DFT functionals are
very inaccurate.

Considering the results on rare-gas systems obtained in the present work and
from previously published papers, the (r)VV10 nonlocal functional seems to
be the most accurate among the DFT methods. It was already shown to be
accurate for many other finite systems,\cite{HujoJCTC11} while for layered
solids it is necessary to modify its parameters (and eventually combine it with
another semilocal functional) to get accurate results.\cite{BjorkmanPRB12} 

\begin{acknowledgments}

The research leading to these results has received funding from the
Swiss University Conference through the High Performance and High
Productivity Computing (HP2C) Programme. Computer resources were
provided by the University of Zurich.

\end{acknowledgments}

\bibliography{references}

\end{document}